# Hole-pocket-driven superconductivity and its universal features in the electron-doped cuprates

Yangmu Li[1,†,*], W. Tabis[1,2], Y. Tang[1], G. Yu[1], J. Jaroszynski[3], N. Barišić[1,4,5,*], and M. Greven[1,*]

[1]School of Physics and Astronomy, University of Minnesota, Minneapolis, Minnesota 55455, USA

[2]Current address: AGH University of Science and Technology, Faculty of Physics and Applied Computer Science, 30-059 Krakow, Poland

[3]National High Magnetic Field National Laboratory, Florida State University, 1800 E. Paul Dirac Drive, Tallahassee, Florida 32310, USA

[4]Institute of Solid State Physics, TU Wien, 1040 Vienna, Austria

[5]Department of Physics, Faculty of Science, University of Zagreb, HR-10000 Zagreb, Croatia

*Correspondence to: yangmuli@umn.edu, neven.barisic@tuwien.ac.at, greven@umn.edu

[†]Current address: Condensed Matter Physics & Materials Science Department, Brookhaven National Laboratory, Upton, NY 11973, USA

Summary: Charge transport measurements reveal hole superconductivity in the electron-doped high-temperature superconductors

**After three decades of enormous scientific inquiry, the emergence of superconductivity in the cuprates remains an unsolved puzzle. One major challenge has been to arrive at a satisfactory understanding of the unusual metallic 'normal state' from which the superconducting state emerges upon cooling. A second challenge has been to achieve a unified understanding of hole- and electron-doped compounds. Here we report detailed magnetoresistivity measurements for the archetypal electron-doped cuprate $Nd_{2-x}Ce_xCuO_{4+\delta}$ that, in combination with prior data, provide crucial links between the normal and superconducting states and between the electron- and hole-doped parts of the phase diagram. The characteristics of the normal state (magnetoresistivity, quantum oscillations, and Hall coefficient) and those of the superconducting state (superfluid density and upper critical field) consistently indicate two-band (electron and hole) features and clearly point to hole-pocket-driven superconductivity in these nominally electron-doped materials. We show that the approximate Uemura scaling between the superconducting transition temperature and the superfluid density found for hole-doped cuprates also holds for the small hole component of the superfluid density in the electron-doped cuprates.**

## INTRODUCTION

Superconductivity in the lamellar cuprates is achieved upon doping the quintessential $CuO_2$ sheets of parent spin-1/2 antiferromagnetic (AF) insulators such as $La_2CuO_4$ and $Nd_2CuO_4$ with either holes[1] or electrons[2]. There has been a resurgence of interest in the electron-doped half of the phase diagram[3-9], where AF correlations are known to be more prominent[10-12]. Hole carriers tend to occupy oxygen $2p$ orbitals, where they frustrate and quickly destroy long-range AF order, whereas electrons primarily enter copper $3d$ orbitals, where they gradually dilute the AF state[10]. In a recent development, normal-state transport measurements revealed Fermi-liquid (FL) properties in a wide temperature and doping range: (1) the sheet resistance follows a FL temperature-doping dependence in the pseudogap regime of the hole-doped cuprates[13-14] and in the AF phase of the electron-doped cuprates[9]; (2) the cotangent of the Hall angle is best understood in terms of a single FL scattering rate that is nearly independent of doping, compound, and charge-carrier type[9,14]; (3) the magnetoresistivity obeys Kohler scaling in the pseudogap regime of the hole-doped materials, with a FL scattering rate[15]; (4) the optical scattering rate exhibits temperature-frequency scaling expected for a FL system[16]. These observations suggest that the transport properties of the cuprates may be understood by FL theory respecting the exact shape of the Fermi surface (FS).

Photoemission[8,17-19] and quantum oscillation[6,7,20] experiments performed on the electron-doped cuprates indicate several distinct FS topologies, as summarized for NCCO in Fig. 1: (1) at low doping, deep in the long-range-ordered AF (LR-AF) phase, only small electron pockets (around $(\pi, 0)$ and equivalent) exist; (2) for bulk superconducting (SC) samples, at intermediate doping, both small electron and hole pockets (around $(\pi/2, \pi/2)$ and equivalent) are observed. Although the AF correlations are short-ranged[12] and dynamic[21,22] in this part of the phase diagram, manifestations of the two-band FS are found in most physical properties. The states (1) and (2) appear to be separated by a "mixed-phase" region, where short-range static AF order and a depressed SC volume fraction are observed, likely as a result of an underlying first-order phase transition in the presence of structural inhomogeneity - e.g., the Nd/Ce substitutional inhomogeneity in the case of $Nd_{2-x}Ce_xCuO_{4+\delta}$ (NCCO) - observed by nuclear magnetic resonance and neutron scattering[23-26]. Additional features of the FS evolution were revealed for a number of electron-doped cuprates, such as NCCO and $Pr_{2-x}Ce_xCuO_{4+\delta}$ (PCCO)[6,27-29]. For example, the Hall coefficient shows a sign change from negative to positive at intermediate doping[6,30] and the Seebeck coefficient has a positive contribution for SC samples[29]. (3) Finally, at high doping, a state with a large hole Fermi surface is expected[17,31], as indeed observed in recent photoemission work[18,19]. The demarcations between these phases depend on the specific compound and choice of annealing conditions. For example, in $La_{2-x}Ce_xCuO_{4+\delta}$ (LCCO) films the boundary between the AF and SC phases has been reported to be as low as $x = 0.07$, yet the nature of the phase transition appears to be the same as in NCCO and PCCO[32].

Dagan and Greene[28] previously studied the planar electrical resistivity (presumably dominated by electrons) and Hall angle (sensitive to both electrons and holes) for PCCO as a function of Ce concentration. They proposed hole superconductivity in the electron-doped cuprates based on the observation that, whereas the resistivity at temperatures much higher than superconducting transition temperature $T_c$ is insensitive to the emergence of superconductivity, the Hall angle can be used to identify the Ce concentration with the highest $T_c$. Yet no direct, quantitative connection was established among the appearance of hole carriers, the normal-state properties, and the SC-state characteristics.

Here we combine new magneto-transport data for NCCO with published results to show that the normal-state properties and the emergence of superconductivity are clearly connected to the shape of Fermi surface and to provide clear evidence for hole-pocket-driven superconductivity in the electron-doped cuprates. Our focus is on the states (1) and (2) and the intermediate mixed-phase region (Fig. 1). The MR magnitude (similar to the Hall angle in ref. 28) is a measure of the overall (hole and electron) FS curvature; it is small for a single, approximately circular FS, but substantial for a two-band (electron-hole) FS (Supplementary Information). We observe a considerable increase in the MR magnitude at the Ce concentration where the small hole pockets and bulk superconductivity are first seen. The normal-state MR therefore reveals the underlying two-band FS topology and the emergent SC ground state. We then perform a quantitative analysis of the electron and hole contributions to the resistive upper critical field and the superfluid density of the bulk SC state. An important early discovery was that the hole-doped compounds below optimal doping exhibit linear scaling between the superfluid density and $T_c$[33]. After separating the two contributions to the superfluid density, we demonstrate that this linear scaling extends to the hole superfluid density of the electron-doped cuprates.

**RESULTS**

Representative transverse *ab*-plane magnetoresistance (MR) data (current $I//a$, magnetic field $H//c$) are shown in Fig. 2. The longitudinal MR ($I//a$, $H//a$) is one order of magnitude smaller than the transverse MR (Fig. S1) and thus not further considered here. The large difference in magnitude implies that the transverse *ab*-plane MR, discussed here in detail, is dominated by orbital contributions. We perform a quantitative analysis of the doping and temperature dependence of the MR using two distinct methods. The MR exhibits quadratic field dependence up to $H_{\text{dev}} \sim 30$ T and $H_{\text{dev}} \sim 15$ T for non-SC and SC samples, respectively. The exact field at which the MR deviates from quadratic field dependence depends on temperature and doping level. We fit the data to $MR \equiv \Delta\rho/\rho(H=0) = b_2 H^2$, where $\Delta\rho \equiv \rho(H) - \rho(H=0)$. $\rho(H=0)$ is the zero-field resistivity at $T > T_c$ and the extrapolated zero-field resistivity at $T \leq T_c$ ("method 1"). The coefficient $b_2$ is a measure of the MR magnitude. The deviation from quadratic field dependence at high fields ($H > H_{dev}$) is largest near optimal doping and indicative of a saturation effect due to the presence of small Fermi pockets. According to the classic theory of magnetoresistivity, the deviation from quadratic field dependence appears when $\omega_c \tau$ becomes larger than unity ($\omega_c$ is the cyclotron frequency and $\tau$ is the relaxation time). The measured $H_{dev} \approx 30$ T for non-SC samples at low doping is consistent with the estimate $H_{dev} \approx 35 \pm 2.5$ T (at $T = 50$ K) based on the reported scattering rate[9] (Supplementary Information). The observation of quantum oscillations requires $\omega_c \tau \gg 1$. At $H \approx 40$ T, quantum oscillations have only been observed in bulk SC samples ($H_{dev} \sim 15$ T)[6,20]. We use the percentage difference, $D_{MR} \equiv (\rho_{H^2} - \Delta\rho)/\rho_{H^2}$, between the extrapolated quadratic behavior, $\rho_{H^2} = b_2 H^2 \rho(H=0)$, and the measured high-field MR, $\Delta\rho$, to characterize the magnitude of the deviation. Because the doping and temperature dependencies of $D_{MR}$ are approximately independent of the field magnitude, we choose $H = 34.5$ T, the highest field used in our experiment, to calculate $D_{MR}$. An alternative approach ("method 2") is to fit the MR over the entire field range to a power-law behavior, $MR = b_n H^n$ (Fig. 2c). In this case, the coefficient $b_n$ is a measure of MR magnitude, whereas $2 - n$ characterizes the magnitude of the deviation from quadratic behavior. Both methods lead to the same conclusion that the normal-state MR and the SC emergence are related to the FS topology. We use method 1 in the main text and compare the two methods in the Supplementary Information.

Contour plots of the coefficient $b_2$ and the high-field deviation $D_{MR}$ are shown in Fig. 1c,d. Both quantities are nearly zero in the LR-AF phase ($x < 0.12$) and substantial for bulk SC materials ($0.145 \leq x < 0.175$), with a distinct increase at about $x = 0.145$. In the mixed-phase region ($0.12 \leq x < 0.145$), $b_2$ is nonzero and clearly related to the SC volume fraction $V_{SC}$ (estimated from magnetization measurements of polycrystalline samples obtained from the crystals used in MR measurements), as demonstrated in Fig.

3a,b. The contour plots of $b_2$ and $D_{MR}$ can be directly compared to the evolution of the FS topology (Fig. 1). Assuming a spatially uniform system, the doping dependence of $b_2$ can be calculated based on the mean-field band structure and Boltzmann theory (Supplementary Information). This calculation indicates that the hole pockets appear due to a decrease of the coherent AF back-scattering amplitude with increasing doping (AF order gives rise to a gap at $(\pi/2, \pi/2)$). Once the small hole pockets are present, a step-like increase is observed in $b_2$. The large values of $b_2$ and $D_{MR}$ reflect a two-band FS in the normal state. In the archetypal cuprate NCCO, mesoscopic phase separation exists in the mixed-phase region, and the dramatic increase of $V_{SC}$ near $x = 0.145$ tracks the doping dependence of $b_2$ (Fig. 3b). As seen from Fig. 3a, the normalized magnetic and superconducting volume fractions approximately add up to unity, which implies that the superconductivity in the mixed-phase region emerges from normal-state regions without static magnetic order.

The electronic ground state of the electron-doped cuprates not only depends on the Ce concentration, but also on the post-growth oxygen reduction conditions[2,3,34-36]. In particular, superconductivity has not been observed in as-grown samples. The phase diagram in Fig. 1 pertains to NCCO subjected to standard oxygen reduction conditions (Methods). Our measurements of NCCO with fixed Ce concentration ($x = 0.170$) show nearly zero MR (single-band FS) for an as-grown, AF sample, and a large MR (two-band FS) for a reduced, bulk-SC sample. A previous study of the oxygen reduction effect on NCCO with $x = 0.15$ also found a strong correlation between the MR and the emergence of superconductivity[30]. These observations suggest a robust connection between the two-band FS and bulk superconductivity, irrespective of the Ce content and oxygen reduction condition. Treating the doping level and reduction condition as implicit parameters, Fig. 3d reveals a quantitative relation between the MR magnitude, normalized by the SC volume fraction, and $T_c$ for a number of thin-film and bulk crystalline NCCO and PCCO samples.

The observation of a distinct signature of the two-band Fermi surface in the MR mandates that other properties should be analyzed accordingly. In particular, the upper critical field ($H_{c2}$) and the superfluid density ($\rho_s$), two characteristics of the SC ground state, may also exhibit two-band features[37,38]. We estimate $H_{c2}$ from our resistivity measurements upon fully suppressing superconductivity at low temperatures, with the magnetic field parallel to the crystalline *c*-axis (Fig. 2b). As demonstrated in Fig. 4a, we observe a universal temperature dependence of the upper critical field for NCCO that is clearly inconsistent with the behavior of a single-band BCS superconductor. This universality implies that disorder effects (Nd/Ce and Cu/Ni substitution, differing oxygen reduction conditions) are not the main cause of the non-single-band temperature dependence of $H_{c2}$. Instead, we find that the data are rather well described by a two-band FS model, analogous to $MgB_2$ and the iron-based superconductors[39,40]. Based on measurements of the Nernst effect, it was argued that the resistive $H_{c2}$ is lower than the "real" (Nernst) upper critical field[41]. We show in the Supplementary Information that $H_{c2}$ estimated from the Nernst effect also cannot be described by a single-band BCS model, but that it can be consistently described by the two-band FS model.

Previous work for the electron-doped cuprates suggested that both electrons and holes contribute to the superfluid density $\rho_s \propto n/m^*$, where $n$ is the normal-state carrier density and $m^*$ is the effective mass[38]. The electron-doped cuprates feature a nonmonotonic SC gap function with nodes at the hole pockets, but not at the electron pockets[42]. The superfluid density for electrons therefore exhibits exponential temperature dependence $\rho_{s,e}(T) = \rho_{s,e}(0)(1 - e^{-\Delta/T + \Delta/T_e})$, where $\rho_{s,e}(0)$ and $\Delta$ are the zero-temperature electron superfluid density and the SC gap at electron pockets, respectively. For simplicity, the latter is assumed to be uniform, as it only changes moderately across electron pockets. $T_e$ is the temperature at which $\rho_{s,e}$ becomes zero and does not necessarily equal $T_c$. Because of existence of the gap node at the hole pockets, in the dirty limit, the superfluid density for holes exhibits a quadratic temperature dependence, $\rho_{s,h}(T) = \rho_{s,h}(0)(1 - T^2/T_h^2)$, where $\rho_{s,h}(0)$ is the zero-temperature hole superfluid density and $T_h$ is the temperature at which the hole superfluid density vanishes. The total superfluid density, $\rho_s(T) = \rho_{s,e}(T) + \rho_{s,h}(T)$, was found to give a good description of prior data[38].

Representative superfluid density data, estimated from transverse-field muon spin rotation/relaxation measurements (µSR) for NCCO ($x = 0.17$), are shown in Fig. 4b. As in ref. 38, we find that $\rho_s(T)$ can only be described by a quadratic-temperature dependence near $T_c$ and by a composite temperature dependence at low temperatures. From the fits, we obtain $\rho_{s,h}(0)/\rho_s(0) = 0.25 \pm 0.03$, and a more conservative estimate yields $\rho_{s,h}(0)/\rho_s(0) = 0.25 \pm 0.06$ (see Methods or Supplementary Information). This value is in reasonably good agreement with quantum oscillation measurements (for $x = 0.15$) that give normal-state electron and hole carrier densities of about 0.18 and 0.03, respectively[20]. Upon considering the different effective masses of electrons and holes ($m^*_{\text{hole}} \approx 0.9 m_e$, $m^*_{\text{electron}} \approx 2 m_e$, where $m_e$ is the electron free mass)[9,14,20], this implies $\rho_{s,h}(0)/\rho_s(0) \approx 0.27$. Similarly, for PCCO, it was reported[38] that $\rho_{s,h}(0)/\rho_s(0) \approx 0.2$. Because data are only available for samples within a close doping range, and hence do not allow for a detailed study of the doping dependence of $\rho_{s,h}(0)/\rho_s(0)$, we assumed the same ratio for each compound (e.g., 0.27 for NCCO and 0.2 for PCCO). In early work, Uemura and colleagues observed a phenomenological universal linear scaling between $\rho_s(0)$ and $T_c$ for underdoped holed-doped cuprates[29]. An approximate linear scaling between these two observables was also found for the electron-doped cuprates (see Fig. 4c), yet the distinct scaling ratios observed for hole- and electron-doped cuprates has remained unexplained[43]. Upon separating electron and hole contributions, we show in Fig. 4c evidence for a universal scaling between $\rho_{s,h}(0)$ and $T_c$ for both electron- and hole-doped cuprates.

**DISCUSSION**

The present MR data together with prior Hall-angle results[28] demonstrate that the emergence of superconductivity can be readily identified via normal-state charge transport measurements at temperatures much higher than $T_c$. This connection extends to other normal-state properties that are sensitive to the balance between hole and electron carrier density, e.g., Hall coefficient[6,30], Seebeck coefficient[29], optical conductivity[31], and quantum oscillations[6,20]. The evolution of the FS topology and the concomitant emergence of hole carriers, as captured by these normal-state charge transport measurements, is responsible for the appearance of a bulk superconducting phase. We note that connections between distinct electronic and structural characteristics and the emergence of superconductivity was reported already in early work[23-26]. MR is a particularly sensitive probe of the emergence of superconductivity, as seen from our results for the mixed-phase region, in which transport properties are the result of a superposition of contributions from AF and SC (non-magnetic) phases. In this region, we find that the MR of NCCO closely tracks the non-magnetic volume fraction (comparable to $V_{sc}$ at low temperature), whereas there has been no report of quantum oscillations, presumably because the typical non-magnetic cluster size is smaller than the characteristic length scale (on the order of the cyclotron radius) associated with the quantum oscillations[7].

As shown in Fig. 4a, we observe a simple scaling of the reduced resistive upper critical field with $T/T_c$ that is well described by a two-band (electron and hole) model. Given that the reduced resistive upper critical field for NCCO is largely independent of disorder type and amount, it is unlikely that the observed temperature dependence is dominated by disorder effects, but rather signifies a universal underlying characteristic of the doped $CuO_2$ planes of the electron-doped cuprates. Raman scattering experiments[44,45] also reveal that the coherent normal-state hole quasiparticles contribute to the superfluid density. Moreover, the superfluid response obtained from penetration depth measurements points to dual electron and hole contributions[38,46]. Upon separating the electron and hole contributions to the superfluid density (Fig. 4b), we find the data in Fig. 4c are consistent with a universal scaling between $\rho_{s,h}(0)$ and $T_c$ for both electron- and hole-doped cuprates. This points to a single underlying hole-related mechanism of superconductivity in the cuprates regardless of nominal carrier type. For the electron-doped cuprates, once the considerable portion of hole pairs have condensed into the SC state, the electrons pairs begin to contribute as well (Fig. 4b).

The carrier density of the $CuO_2$ planes, and hence the Fermi surface of the electron-doped cuprates can be modified by two methods: (1) chemical substitution (nominally tetra-valent Ce for tri-valent La, Nd, Pr, or Sm) and (2) a post-growth oxygen reduction process; both methods alter the disorder potential experienced by the $CuO_2$ planes[2,3,8,34-36]. Our $x = 0.17$ NCCO MR data (Fig. 3c) along with the prior $x = 0.15$ NCCO result[30] show that a correlation between $T_c$ and the normal-state MR exists in both cases. Recently, superconductivity was achieved in Ce-free thin-film samples via a special reduction procedure[3]. Remarkably, both the Fermi surface revealed by quantum oscillation measurements (the existence of small hole pockets) and the MR ($b_2 \sim 1.6 \times 10^{-4}$) for Ce-free thin-film samples of $Pr_2CuO_{4\pm\delta}$ are the same as that for the Ce-doped bulk-superconducting samples subjected to standard oxygen reduction[7]. These FS characteristics imply that these Ce-free SC thin-film samples are not undoped, but instead correspond to region (2) of the phase diagram (Fig. 1). Moreover, as shown in Fig. 4c, the estimated hole superfluid density obeys the universal scaling established here, indicative of the same SC ground state irrespective of reduction conditions.

The infinite-layer cuprates $Sr_{1-x}Ln_xCuO_2$ (Ln = La, Nd, Pr, Sm) constitute a second family of electron-doped materials that differs structurally from the T′ family $Ln_{2-x}Ce_xCuO_4$ (Ln = La, Nd, Pr, Sm, Eu and Gd). The exact symmetry of the superconducting wave function is under debate (Supplementary Information) and the possible emergence of superconductivity from electron Fermi pockets was reported[47,48]. In light of the fact that a hole contribution was deduced from the normal-state Hall constant of superconducting samples[49], we also considered the superfluid density of the infinite-layer cuprates. One way to understand the superfluid density of $Sr_{0.9}La_{0.1}CuO_2$ is to decompose it into *s*-wave and *d*-wave contributions[47]. Assuming that the *s*-wave contribution is due to electrons and the *d*-wave contribution due to holes, we show in Fig. 4c that the universal scaling seems to be obeyed as well.

## MATERIALS AND METHODS

### Sample preparation

The NCCO samples were synthesized using the traveling-solvent floating-zone technique in 4 atm Ar/O$_2$ atmosphere, and oriented by Laue diffraction within an angle of ± 2 degrees. Samples were reduced for 12 hours at 970°C in Ar flow and then treated for 20 hours at 500°C in oxygen flow. The onset superconducting transition temperature was determined from magnetization measurements using a Quantum Design, Inc., Magnetic Property Measurement System (MPMS) and from resistivity measurements. The Ce concentration was measured with inductively-coupled plasma atomic emission spectroscopy and/or energy-dispersive X-ray spectroscopy. Approximately 1 mm of the surface of each crystal was removed to improve Ce homogeneity.

### MR measurements and analysis

Single crystals of NCCO with $x = 0.110(10), 0.133(3), 0.145(4), 0.156(4), 0.170(2)$ were measured using the four-contact method or in the Hall-bar configuration. Measurements were performed with a Quantum Design, Inc., Physical Properties Measurement System (PPMS, up to 9 Tesla) and at National High Magnetic Field Laboratory (dc field up to 34.5 Tesla). The MR was determined in two principal geometries (I//a, H//a; I//a, H//c). For a few samples [$x = 0.133(3), 0.145(4), 0.156(4)$], the angular dependence was obtained. For simplicity, $\mu_0$ and $k_B$ are set to 1 throughout this work. For additional details, see Supplementary Information.

**Acknowledgements:** We thank E. M. Motoyama for the growth of some of the NCCO samples, I. M. Vishik for the growth of the Ni-doped NCCO sample studied in this work, T. Peterson for assistance with the operation of the Quantum Design, Inc., PPMS system, and J. Cai for help with the MR calculations. We gratefully acknowledge helpful discussions with A. V. Chubukov, R. L. Greene, J. Kang, E. H. da Silva Neto, I. M Vishik and X. Wang. This work was supported partially by the NSF through the University of Minnesota MRSEC under Grant No. DMR-1420013 and by NSF Grant No. 1006617. The work at the TU Wien was supported by FWF project P27980 - N36 and the European Research Council (ERC Consolidator Grant No 725521). A portion of this work was performed at the National High Magnetic Field Laboratory, which is supported by National Science Foundation Cooperative Agreement No. DMR-1157490 and the State of Florida.

**Author contributions:** Y.L., N.B., M.G. conceived the research; Y.L., W.T., G.Y. prepared the samples for transport measurements; Y.L., N.B., W.T., Y.T., and J. J. carried out the measurements at the NMHFL; Y.L. carried out in-house magnetoresistivity measurements; Y.L. performed data analysis and calculations guided by N.B and M.G.; Y.L., N.B., M.G. wrote the manuscript with input from all authors.

**Competing interests:** The authors declare that they have no competing interests.

**Data and materials availability:** All data needed to evaluate the conclusions in the paper are present in the paper and/or the Supplementary Materials. Additional data related to this paper may be requested from the authors.

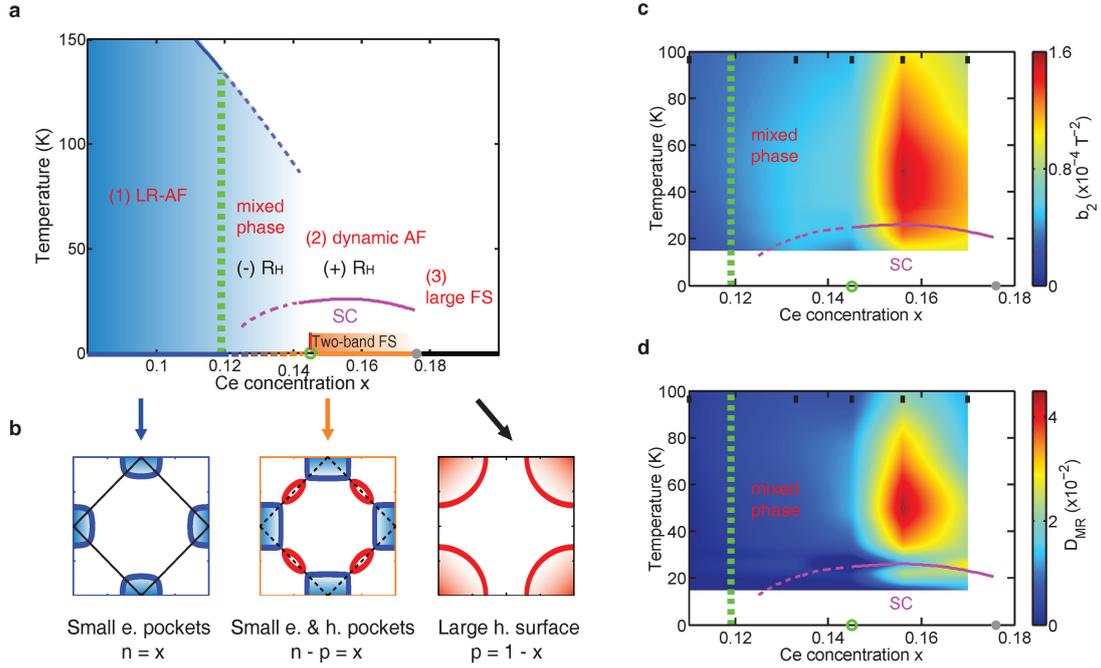

**Fig 1. Phase diagram and Fermi-surface topologies. (a)** Magnetic and electronic phase diagram for NCCO with standard reduction condition. The vertical green dashed line shows the boundary between the (1) long-range antiferromagnetic (LR-AF) phase ($x < 0.12$) and the "mixed" phase region with static short-range antiferromagnetic order and traces of superconductivity ($0.12 \leq x < 0.145$). The empty green circle signifies the Ce concentration at which (2) bulk superconductivity emerges and the magnetic response is purely dynamic ($0.145 \leq x < 0.175$)[22]. A sign change in Hall coefficient is observed at $x = 0.145$[6]. The grey dot shows the estimated Ce concentration of the Lifshitz transition to a (3) state with a large hole FS ($x \geq 0.175$)[6,17,18,20,31]. Solid and dashed lines on horizontal axis indicate distinct FS topologies described in (b). **(b)** FS topologies corresponding to the three doping ranges in (a). Solid ($x < 0.12$) and dashed ($0.145 \leq x < 0.175$) diagonal black lines indicate the LR-AF zone boundary and dynamic AF fluctuations, respectively. Blue and red curves indicate electron and hole FS. $n$ and $p$ are electron and hole carrier densities, respectively. Contour plot of the MR **(c)** $b_2$ and **(d)** $D_{MR}$, respectively. A considerable increase both in the $b_2$ and $D_{MR}$ is observed at $x \approx 0.145$. The black bars (top) indicate the Ce concentrations of the measured NCCO samples. Color scheme of the contour plots is chosen to emphasize such considerable increases.

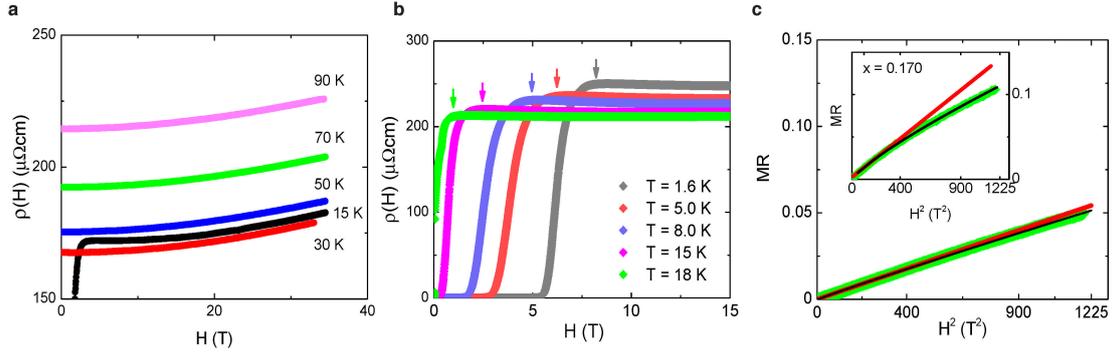

**Fig 2. Representative magnetoresistivity data.** (a) Raw magnetoresistivity for NCCO with $x = 0.133$ and $0.170$. For the $x = 0.133$ sample, a weak resistivity upturn in temperature can be seen below 30 K[9]. (b) Magnetoresistivity at $T < T_c$ for $x = 0.133$. The arrows indicate the field at which the SC is fully suppressed and the normal-state resistivity is recovered (resistive upper critical field). Note that for non-SC samples at low doping ($x \leq 0.11$), the magnetoresistivity can be negative at low temperatures. (c) MR (green) for $x = 0.133$ and $x = 0.170$ (inset) analyzed by two methods, as described in the text ("method 1": red lines; "method 2": black curves).

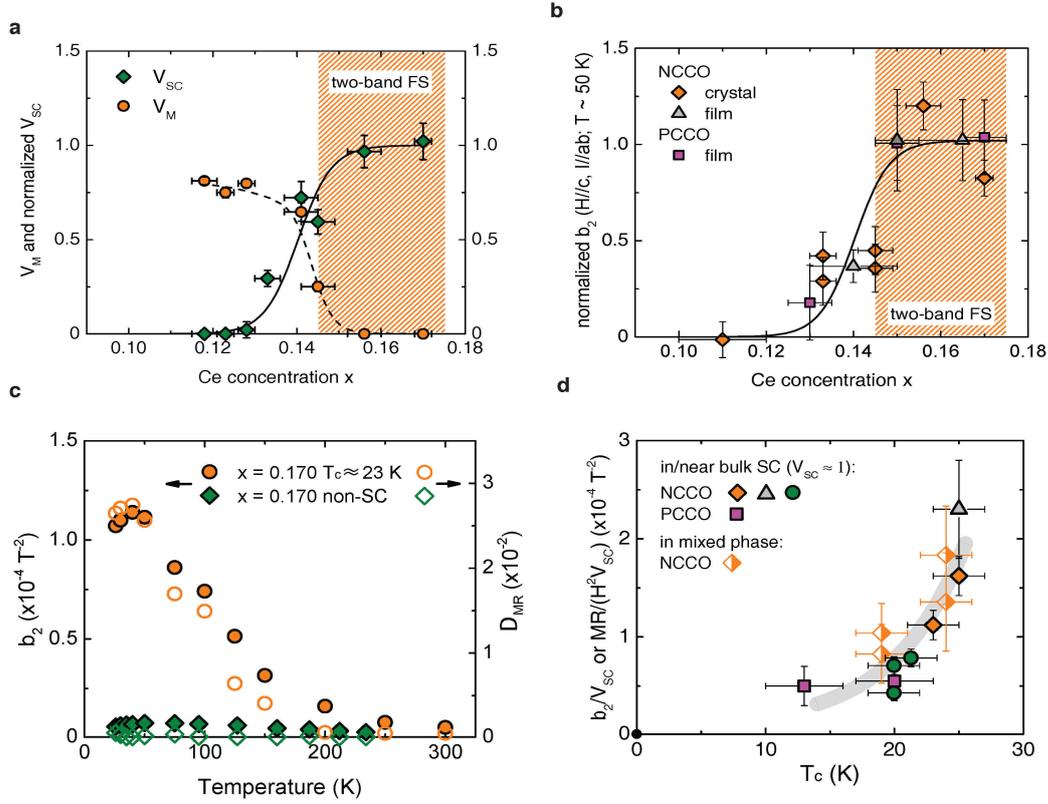

**Fig 3. Correlation between magnetoresistivity and the emergence of superconductivity.** (a) SC and magnetic volume fractions as a function of Ce concentration. The former was obtained from magnetization data at 2 K whereas the latter was obtained from μSR measurements at 40 K. As a result of doping the CuO$_2$ planes with electrons, the magnetic volume fraction has decreased to $V_M \approx 0.82$ at $x \approx 0.12$. Dashed and solid lines are guides to the eye. Shaded area indicates the doping range of the two-band FS. Reproduced from ref. 22. (b) Normalized MR coefficient $b_2$ ($b_2$ normalized to 1 for samples with $x \approx 0.15$) as function of Ce concentration. The symbol labels are the same as in **d**. (c) Coefficients $b_2$ and $D_{MR}$ for as-grown, non-SC and reduced, SC NCCO with $x = 0.170$. (d) MR normalized by SC volume fraction (non-magnetic volume fraction in the normal state) for NCCO and PCCO films and crystals. Orange diamonds: present work for NCCO. Grey triangles[50] and Green circles[30]: NCCO films. Purple squares: PCCO films[27]. All the MR data in this figure were taken at $T \approx 50$ K. The full symbols indicate samples for which $V_{SC} \approx 1$ is estimated. The half symbols signify samples in the "mixed" phase. Error bars show estimated one standard deviation.

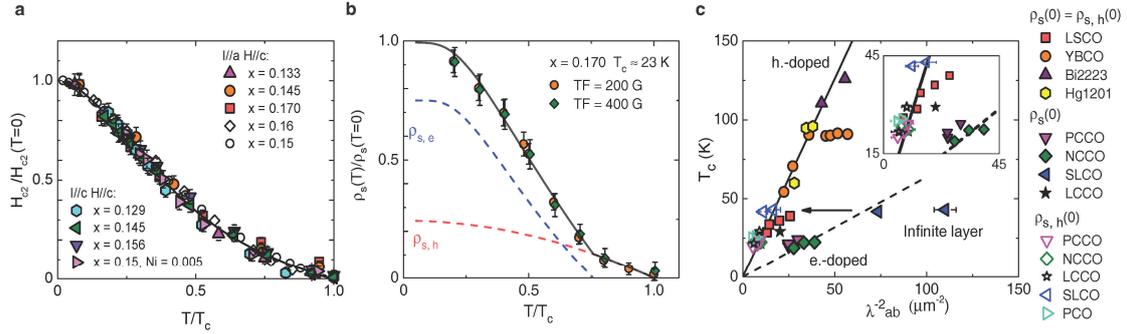

**Fig 4. Two-band upper critical field and superfluid density.** (**a**) Universal temperature dependence of the resistive upper critical field for NCCO with *H*//c. The solid curve is a fit to the two-band model (Supplementary Information). Published data taken from refs. 37 ($x = 0.15$) and 50 ($x = 0.16$). (**b**) Temperature dependence of superfluid density extracted from the μSR measurements for a $x = 0.170$ sample[22]. Fit of the two-band model, as described in the text, is shown as a black solid line. Electron and hole contributions are plotted as blue and red dashed lines, respectively. (**c**) Universal scaling between $\rho_{s,h}(0)$ and $T_c$ (black solid line). Black dashed line present previous scaling[33] between $\rho_s(0)$ and $T_c$ for the electron-doped cuprates. Both $\rho_{s,h}(0)$ and $\rho_s(0)$ are plotted as the inverse square of the penetration depth $\lambda_{ab}$. Data adopted from ref. 22 (NCCO), ref. 43 (PCCO, NCCO, SLCO), ref. 47 (SLCO), ref. 51 (La$_{2-x}$Sr$_x$CuO$_4$ (LSCO), YBa$_2$Cu$_3$O$_y$ (YBCO), Bi$_2$Sr$_2$Ca$_2$Cu$_3$O$_{10+\delta}$ (Bi2223)), ref. 52 (HgBa$_2$CuO$_{4+\delta}$ (Hg1201)), and ref. 53 (LCCO). The superfluid density of Ce-free Pr$_2$CuO$_{4\pm\delta}$ (PCO) samples subjected to special reduction conditions is estimated based on measurements reported in ref. 14. Note that the superfluid density of samples in the mixed phase can be somewhat smaller than the scaling result due to phase separation.

**List of Supplementary Materials**

Sample preparation

Table S1. NCCO sample information.

Supplementary results and discussion

1. Magnetoresistivity

2. Magnetoresistivity for H//a

3. Estimation of magnetoresistivity

4. Resistive and Nernst upper critical fields

5. Superfluid density

Fig S1. Longitudinal (I//a, H//a) ab-plane magnetoresistivity.

Fig S2. Comparison between two methods to analyze magnetoresistivity.

Fig S3. Fermi surface and calculation of magnetoresistivity.

Fig S4. Nernst upper critical field.

Fig S5. Representative MR and $H_{C2}$ data.

# Supplementary Materials for

## "Hole-pocket-driven superconductivity and its universal features in the electron-doped cuprates"

Yangmu Li[1,†,*], W. Tabis[1,2], Y. Tang[1], G. Yu[1], J. Jaroszynski[3], N. Barišić[1,4,5,*], and M. Greven[1,*]

[1]School of Physics and Astronomy, University of Minnesota, Minneapolis, Minnesota 55455, USA

[2]Current address: AGH University of Science and Technology, Faculty of Physics and Applied Computer Science, 30-059 Krakow, Poland

[3]National High Magnetic Field National Laboratory, Florida State University, 1800 E. Paul Dirac Drive, Tallahassee, Florida 32310, USA

[4]Institute of Solid State Physics, TU Wien, 1040 Vienna, Austria

[5]Department of Physics, Faculty of Science, University of Zagreb, HR-10000 Zagreb, Croatia

*Correspondence to: yangmuli@umn.edu, neven.barisic@tuwien.ac.at, greven@umn.edu

†Current address: Condensed Matter Physics & Materials Science Department, Brookhaven National Laboratory, Upton, NY 11973, USA

**List of Supplementary Materials**

Sample preparation

Table S1. NCCO sample information.

Supplementary results and discussion

1. Magnetoresistivity

2. Magnetoresistivity for H//a

3. Estimation of magnetoresistivity

4. Resistive and Nernst upper critical fields

5. Superfluid density

Fig S1. Longitudinal (I//a, H//a) ab-plane magnetoresistivity.

Fig S2. Comparison between two methods to analyze magnetoresistivity.

Fig S3. Fermi surface and calculation of magnetoresistivity.

Fig S4. Nernst upper critical field.

Fig S5. Representative MR and $H_{C2}$ data.

**Sample preparation**

The NCCO crystals measured in this work were characterized by inductively coupled plasma atomic emission spectroscopy (ICP) and/or energy-dispersive X-ray spectroscopy (EDS). Some of the crystals were previously measured using neutron scattering and µSR techniques[11,12,22]. Because a difference in Ce concentration was observed between sample surface and interior, the cylindrical crystals were polished. This led to a typical decrease of the crystal radius of about 1 mm and to improved overall homogeneity. Samples for transport measurements were cut from big pieces of single crystals.

|  | Nominal x | Actual x | $T_c$ (K) | Comment | I//a, H//c | I//a, H//a |
|---|---|---|---|---|---|---|
| Z10 | 0.10 | 0.110(10) | 0 |  | X |  |
| E19[‡] | 0.125 | 0.133(3) | 24 |  | X | X |
| E9[†§] | 0.1375 | 0.145(4) | 19 |  | X | X |
| E23[‡§] | 0.146 | 0.156(4) | 25 |  | X | X |
| YM16[§] | 0.16 | 0.170(2) | 23 |  | X |  |
| YM16 (AG)[§] | 0.16 | 0.170(2) | 0 | as-grown | X |  |

**Table S1. NCCO sample information.**

X: Magnetoresistivity measured with magnetic field up to $H = 34.5$ T at the National High Magnetic Field Laboratory and /or via a Quantum Design Inc. Physical Properties Measurement System (up to $H = 9$ T).

[†]: Measured by neutron scattering in ref. 12. [‡]: Measured by neutron scattering in ref. 22. [§] Measured by µSR in ref. 22.

**Supplementary results and discussion**

**1. Magnetoresistivity**

Magnetoresistivity has been extensively used to study the Fermi surface (FS) of various materials. At very high magnetic fields ($\omega_c \tau \gg 1$, where $\omega_c$ is the synchrotron frequency and $\tau$ is the scattering mean free time), quantum oscillations (Shubnikov-de Haas oscillations) occur as the Landau levels pass through the Fermi energy (e. g., ref. 6). The oscillation frequency (with regard to the inverse magnetic field strength) is a measure of the size of high-field FS. At low field ($\omega_c \tau \ll 1$), where FS reconstruction induced by the magnetic field is of much less concern, $MR \equiv \Delta\rho/\rho(H=0) \propto b_2 H^2$ is expected for a simple two-band system with parabolic dispersion, where $\rho(H=0)$ is the resistivity at zero magnetic field and $\Delta\rho = \rho(H) - \rho(H=0)$. In low magnetic fields, the magnitude of the magnetoresistivity ($b_2$) of an open FS is much less than that of a closed FS. At intermediate field strength ($\omega_c \tau \approx 1$), in the case of a closed FS, the magnetoresistivity saturates and deviates from the quadratic field dependence, whereas that of an open Fermi surfaces continues to behave as $MR \propto b_2 H^2$ and become much larger. For NCCO, at intermediate doping levels, both electron and hole Fermi pockets exist. The size of hole pockets is much smaller than that of the electron pockets. The deviation of the magnetoresistivity from quadratic field dependence is thus expected to be caused by a saturation effect associated with the small hole pockets.

The magnetic field at which the saturation effect appears ($H_{\text{dev}}$) can be estimated using the scattering rate $\tau$. The scattering rate in the underdoped regime for the electron-doped cuprates was reported to be $\tau = (m^*/ecT^2)$, where $m^*$ is the effective mass, $e$ is the electron charge, $T$ is temperature, and $c$ is a constant[9]. For $x = 0.10$ NCCO, $c = 0.014 \pm 0.001$. Therefore, from

$$1 = \omega_c \tau = \left(\frac{eH_{\text{dev}}}{m^*}\right)\left(\frac{m^*}{ecT^2}\right),$$

we estimate $H_{\text{dev}} = 35 \pm 2.5$ T.

**2. Magnetoresistivity for H//a**

The magnetoresistivity for H//a shows no apparent deviation from $MR \propto b_2 H^2$ up to the highest measured magnetic field, and the values of $b_2$ for H//a are much smaller than those for H//c. In Fig. S1, we compare the two and, show representative data for the angular dependence of the magnetoresistivity, as well as contour plots of $b_2$ and $D_{MR}$ for H//a (similar to Fig. 3 in the main text). The latter feature no discernible doping dependences. These results suggest the existence of a two-dimensional FS.

## 3. Estimation of magnetoresistivity

In Fig. S2, we show similar results obtained using the two methods to analyze magnetoresistivity (see main text). The estimation of magnetoresistivity based on FS topology is described in the following.

The FS is calculated using tight-binding band structure based on the local density approximation[54, 55].

The dispersions are $E_k^\pm = 1/2[\varepsilon_k + \varepsilon_{k+(\pi,\pi)} \pm \sqrt{(\varepsilon_k - \varepsilon_{k+(\pi,\pi)})^2 + 4\Delta^2}$, where $\varepsilon_k$ and $\varepsilon_{k+(\pi,\pi)}$ are the original band dispersion in the absence of AF order and the band dispersion shifted by the AF propagation vector, respectively, and $\varepsilon_k = -2t_1(\cos k_x + \cos k_y) + 4t_2 \cos k_x \cos k_y - 2t_3(\cos 2k_x + \cos 2k_y) + \mu$. $\Delta$ is the coherent back scattering amplitude (or the band gap at $(\pi/2, \pi/2)$), and $\mu$ is the chemical potential. Note that both chemical substitution and oxygen reduction change $\Delta$. For the purpose of estimating the effect of changes in the Fermi surface topology on the magnetoresistivity, we do not consider any dependence of band parameters on chemical substitution. The band parameters used are $t_1 = 0.38$ eV, $t_2 = 0.3 t_1$ and $t_3 = 0.18 t_1$. The chemical potential is kept at zero for simplicity. For each value of $\Delta$, we performed a numerical integration of the FS area to determine the electron ($n_e$) and hole ($n_h$) carrier densities, with the constraint $n_e - n_h = x$, where $x$ is the Ce concentration. The result are shown in Fig. S3.

For the estimation of the magnetoresistivity, three methods were used:

(1) Simple two-band model: for two types of carriers with densities $n_e$ and $n_h$ and mobilities $\mu_e$ and $\mu_h$, the transverse magnetoresistivity[56] can be expressed as $MR = [n_e n_h \mu_e \mu_h (\mu_e - \mu_h)^2] H^2 / (n_e \mu_e + n_h \mu_h)^2 = n_e n_h \mu_e \mu_h H^2 (\tau_e/m_e^* - \tau_h/m_h^*)^2 / (n_e \tau_e / m_e^* + n_h \tau_h / m_h^*)^2$, where $\tau_e$, $\tau_h$, $m_e^*$ and $m_h^*$ are electron relaxation time, hole relaxation time, electron effective mass and hole effective mass, respectively. Recent transport studies[9,14] suggest that both $\tau_e / m_e^*$ and $\tau_h / m_h^*$ are proportional to $T^{-2}$. We thus assume $\tau_e / m_e^* = \alpha \tau_h / m_h^*$, where $\alpha$ is a constant ($\alpha \neq 1$). At a qualitative level, the result does not depend on the exact value of $\alpha$. As can be seen, $MR = 0$ when $n_h = 0$.

(2) Boltzmann + FS details: calculations using Boltzmann theory on a two-dimensional, fourfold symmetric, but otherwise arbitrary FS found that the magnetoresistivity depends on the details of the Fermi surface curvature and $MR = (|e|\mu_0 H/\hbar)^2 [\langle (dl/ds)^2 \rangle_\Sigma + \langle (ld\vartheta/ds) \rangle_\Sigma^2 - \langle (ld\vartheta/ds)^2 \rangle_\Sigma]$, where $l$, $s$ and $\vartheta$ are the local mean free path, arc length along the Fermi Surface and the angle between velocity of the carriers and the electrical field. The averaging $\langle ... \rangle_\Sigma$ means $\int \Sigma(s) ... ds$, where $\Sigma(s)$ is the conductivity weight of $s$, and the $d\vartheta/ds$ is the local curvature of the FS[50]. Here, we calculate the MR numerically.

According to ref. 50, $\langle (dl(\mathbf{k})/ds)^2 \rangle_\Sigma = \int (\tau(\mathbf{k})dv(\mathbf{k})/ds)^2 \sigma(\mathbf{k})ds / \int \sigma(\mathbf{k})ds$, where $\tau(\mathbf{k})$, $v(\mathbf{k})$ and $\sigma(\mathbf{k})$ is the local scattering relaxation time, local Fermi velocity and local conductivity, respectively. $s$ is a function of $\mathbf{k}$. We assume an isotropic relaxation time $\tau$. $v(\mathbf{k}) = dE_k^+/d\mathbf{k}$ for hole pockets and $v(\mathbf{k}) = dE_k^-/d\mathbf{k}$ for electron pockets.

$\sigma(\mathbf{k}) = e^2 l(\mathbf{k})/(4\pi^2 \hbar) = e^2 \tau v(\mathbf{k})/(4\pi^2 \hbar)$,

$\langle (ld\vartheta/ds) \rangle_\Sigma^2 = (\int (\tau v(\mathbf{k})d\vartheta/ds)\sigma(\mathbf{k})ds / \int \sigma(\mathbf{k})ds)^2$,

and $\langle (ld\vartheta/ds)^2 \rangle_\Sigma = (\int (\tau v(\mathbf{k})d\vartheta/ds)\sigma(\mathbf{k})ds)^2 / \int \sigma(\mathbf{k})ds$.

The derivative $d\vartheta/ds$ is positive for electron and negative for hole pockets.

(3) Spectral function: the spectral function is calculated such that[57]

$$A_k^\pm(\omega) = \frac{1}{\pi} \frac{W_k^\pm \Gamma^\pm}{(\omega - E_k^\pm)^2 - (\Gamma^\pm)^2}$$

Where $\Gamma^\pm$ is the scattering rate (assuming to be isotropic) and $W_k^\pm$ is the weighting function

$$W_k^\pm = \frac{1}{2}\left[1 \pm \frac{\varepsilon_k + \varepsilon_{k+(\pi,\pi)}}{\sqrt{(\varepsilon_k - \varepsilon_{k+(\pi,\pi)})^2 + 4\Delta^2}}\right]$$

The conductivity is:

$$\sigma_{xx}^\pm = \sigma_{yy}^\pm \propto \int d\omega \sum_k \left[\frac{\partial E_k^\pm}{\partial k_x}\right]^2 [A_k^\pm(\omega)]^2$$

$$\sigma_{xy}^\pm \propto \int d\omega \sum_k \frac{\partial E_k^\pm}{\partial k_x}\left[\frac{\partial E_k^\pm}{\partial k_x}\frac{\partial^2 E_k^\pm}{\partial k_y^2} - \frac{\partial E_k^\pm}{\partial k_y}\frac{\partial^2 E_k^\pm}{\partial k_x^2}\right] [A_k^\pm(\omega)]^3$$

The magnetoresistivity is:

$$MR_{xx} = \mu_0^2 H^2 \sigma_{xx}^+ \sigma_{xx}^- (\sigma_{xy}^+ - \sigma_{xy}^-)^2 / (\sigma_{xx}^+ + \sigma_{xx}^-)^2$$

Using numerical summation, we obtain the $MR_{xx}$ based on the FS and band structure given by the local density approximation. This method is similar to method (2).

Note that for all methods we do not consider a "mixed" phase region. Ref. 58, using method (2), gave the temperature dependence of the MR in good agreement with the experiment. We do not repeat the calculation of the temperature dependence here.

## 4. Resistive and Nernst upper critical fields

Following refs. 39 and 40, the resistive upper critical field $H_{c2}$ is described by the parametric equation:

$$\ln\left(\frac{T_c}{T}\right) = \frac{1}{2}\left[U(s) + U(\eta s) + \frac{\lambda_0}{w}\right] - \left\{\frac{1}{4}\left[U(s) - U(\eta s) - \frac{\lambda_-}{w}\right]^2 + \frac{\lambda_{eh}\lambda_{he}}{w^2}\right\}^{1/2}$$

$$H_{C2} = \frac{2\phi_0 T s}{D_e} \qquad \eta = D_h/D_e$$

$$U(s) = \psi\left(s + \frac{1}{2}\right) - \psi\left(\frac{1}{2}\right)$$

where $\lambda_- = \lambda_{ee} - \lambda_{hh}$, $\lambda_0 = (\lambda_-^2 + 4\lambda_{eh}\lambda_{he})^{1/2}$, and $w = \lambda_{ee}\lambda_{hh} - \lambda_{eh}\lambda_{he}$, $\lambda_{ee}$, $\lambda_{hh}$, $\lambda_{eh}$, $\lambda_{he}$ are matrix elements of the BCS coupling constants. $D_e$ and $D_h$ are the electron and hole diffusivities. $\phi_0$ is the magnetic flux quantum. $\psi(x)$ is the digamma function. In the fit of $H_{c2}$ in Fig. 4a, we obtain $D_e \approx 0.885 \pm 0.139$, $D_h \approx 0.051 \pm 0.003$, $\lambda_{ee} \approx 0.250 \pm 0.019$, $\lambda_{hh} \approx 0.650 \pm 0.036$ and $\lambda_{eh} = \lambda_{eh} = 0$. Note that in this case the upper critical field is mainly determined by the $D_h$[39]. We carried out both decoupled-band ($\lambda_{eh} = \lambda_{he} = 0$) and coupled-band (with free parameters $\lambda_{eh}$ and $\lambda_{he}$) fits in the analysis of the upper critical field. In all cases, the coupled-band fit resulted in negligible values of $\lambda_{eh}$ and $\lambda_{he}$.

Based on measurements of the Nernst effect, it was argued that $H_{c2}$ estimated from the resistivity corresponds to the field at which the Nernst signal is maximum ($H^*$), and hence that it is lower than the actual upper critical field[41]. We note that $H^*$ is the field at which the resistivity becomes nonzero (once the vortex solid melts), and that it is smaller than the resistive upper critical field at which resistivity recovers to its normal-state value (our definition)[59]. $H_{c2}$ determined from the Nernst effect or deduced from $H^*$ were argued to be less reliable in the case of cuprates[59]. Our analysis of the prior Nernst data[41] reveals that, although a simple single-band BCS model does not describe the temperature dependence of the Nernst upper critical field, a two-band model describes data rather well (see Fig. S4). The fit resulted in the following parameters: $D_e \approx 0.761 \pm 0.475$, $D_h \approx 0.102 \pm 0.014$, $\lambda_{ee} \approx 0.151 \pm 0.020$, and $\lambda_{hh} \approx 0.857 \pm 0.095$ (with $\lambda_{eh} = \lambda_{he} = 0$). We applied both decoupled-band ($\lambda_{eh} = \lambda_{he} = 0$) and coupled-band free fit ($\lambda_{eh}$, $\lambda_{he}$ are free parameters) conditions in the analysis of the upper critical field. In all cases, we found $\lambda_{eh}$ and $\lambda_{he}$ are negligible small. The fit results do not depend on these conditions.

The parameters determined based on both resistive and Nernst upper critical fields show $\lambda_{hh} \gg \lambda_{ee}$, and $\lambda_{eh} \approx \lambda_{eh} \approx 0$, indicative of hole-pocket-driven superconductivity.

**5. Superfluid density**

The fit of the superfluid density data for $x = 0.17$ NCCO (Fig. 4b) is based on the two-band model described in the main text and follows the following logic: (1) at high temperatures ($T \sim T_c$), the superfluid density is best described by the hole contribution $\rho_s \sim (1 - T^2/T_c^2)$, and we can set $T_h = T_c$; (2) at low temperatures, both electron and hole contributions are required to fit the total superfluid density, and $T_e$ is determined by the fit. Our analysis gives $T_e = (0.75 \pm 0.05)\,T_c$. We also attempted to fit the data with a fixed $T_e = T_c$, which resulted in a much worse fit. The result $T_h = T_c > T_e$ is indicative of hole-pocket-driven superconductivity. We estimate that the average gap for electrons is $2\Delta_{avg} = 5 \pm 1$ meV and maximum gap is $2\Delta_{max} = 6.7 \pm 1.3$ meV, consistent with Raman spectroscopy[44].

The errorbar of $\rho_{s,h}(0)/\rho_s(0) = 0.25 \pm 0.03$ comes from the result of the fit with the error for each datum is taken into consideration. In order to further demonstrate the robustness of the result, the error estimation is pushed to the extreme. A more conservative errorbar is obtained by intentionally fit $\rho_{s,h}(0)$ minus its one standard deviation for all data points with $T > T_e$ and $\rho_{s,h}(0)$ plus its one standard deviation for all data points with $T < T_e$. This resulted in $\rho_{s,h}(0)/\rho_s(0) = 0.25 \pm 0.06$.

Superconductivity was reported along with the observation of the small hole pockets in nominally undoped PCO films subjected to a special oxygen reduction process[3,7,8]. The hole-carrier density can be estimated from the oscillation frequency $F$. Using the values $F = 290$ T for NCCO[20] ($T_c \approx 25$ K, $x = 0.15$) and $F = 310$ T for the specially reduced PCO films[7] ($T_c \approx 26$ K) and the corresponding effective masses from quantum oscillation[7], we show in Fig. 4c that nominally undoped superconducting PCO obeys the universal scaling relationship between the hole superfluid density and $T_c$. Here we assumed that all the normal-state hole carriers contribute to the superfluid density.

The infinite-layer cuprates $Sr_{1-x}Ln_xCuO_2$ (Ln = La, Nd, Pr, Sm) constitute a second family of electron-doped materials. The exact symmetry of their superconducting wave function is under debate. An *s*-wave contribution was reported[47,48], and it was also argued that the superconductivity is observed in a regime where only electron pockets exist[48]. Although photoemission measurements revealed intensity near the putative hole pocket positions, this was argued to be due to the tail of the broad quasiparticle spectral function rather than hole carriers. In ref. 46, the superfluid density was analyzed and argued to be consistent with a decomposition into 85% *s*-wave and a 15% *d*-wave contributions. We note that is only one of the possible ways to understand the superfluid density. Similar to the T′ family $Ln_{2-x}Ce_xCuO_4$ (Ln = La, Nd, Pr, Sm, Eu and Gd), the Hall constant[49] for the infinite-layer compounds indicates a hole contribution in the normal state of superconducting samples. In our analysis summarized in Fig. 4c, the electron contribution is described with an isotropic (*s*-wave) superconducting gap, and the hole contribution is described using a superconducting gap with a node (*d*-wave). If assume that the 85% *s*-wave and the 15% *d*-wave contributions to the superfluid density in $Sr_{1-x}Ln_xCuO_2$ are due to electrons and holes, respectively, we arrive at the relationship between hole superfluid density and $T_c$ shown in Fig. 4c.

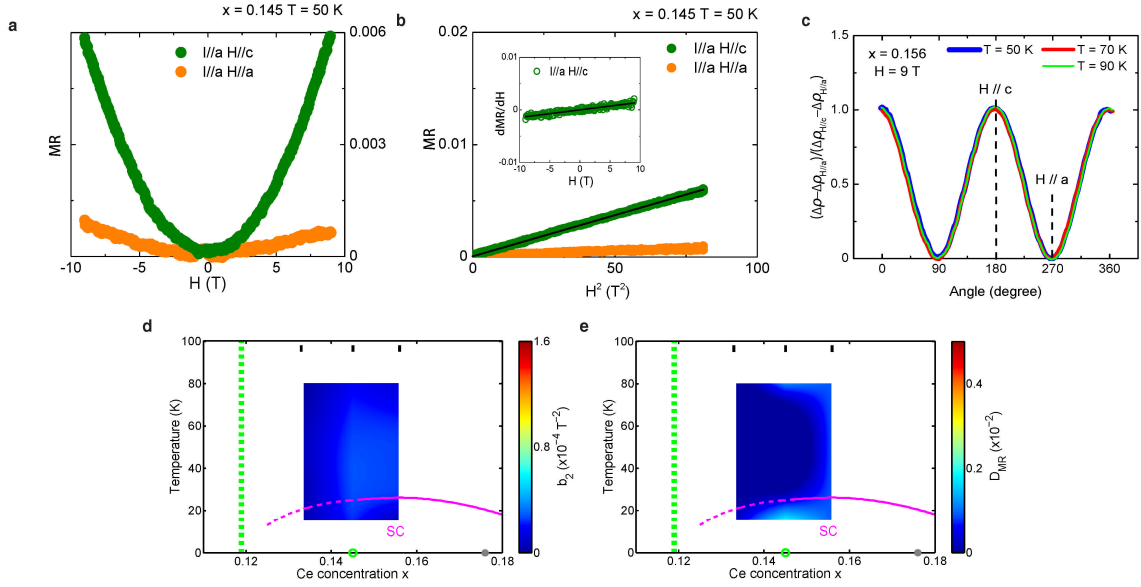

**Fig S1. Longitudinal (I//a, H//a) ab-plane magnetoresistivity. (a, b)** Comparison between longitudinal (I//a, H//a) and transverse (I//a, H//c) *ab*-plane MR. Both MR have quadratic field dependence at low magnetic field. The longitudinal (I//a, H//a) MR is one order of magnitude smaller than the transverse (I//a, H//c) MR. The inset of (b) shows the first-order numerical derivative of the transverse MR, demonstrating the quadratic field dependence. **(c)** Representative angular dependence of the MR. **(d, e)** Contour plots of $b_2$ and $D_{MR}$ for the longitudinal MR. All labels are the same as in Fig. 3 of the main text.

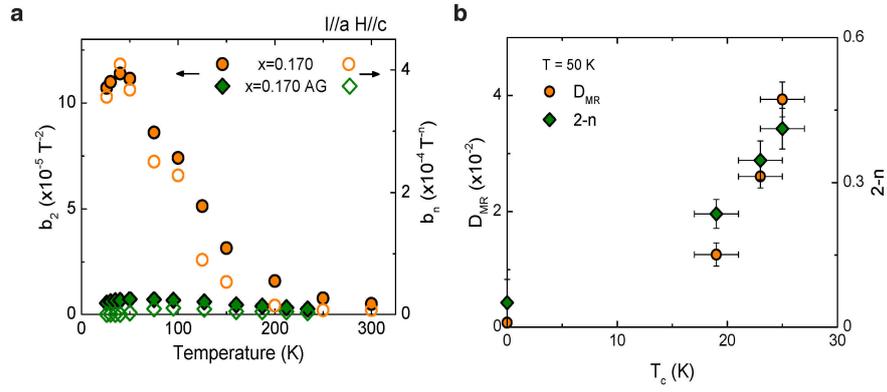

**Fig S2. Comparison between two methods to analyze magnetoresistivity.** (a) Comparison between $b_2$ and $b_n$ of reduced and as-grown $x = 0.170$ NCCO samples. The coefficients $b_2$ and $b_n$ have similar temperature dependences. (b) Comparison between $D_{MR}$ and $2 - n$. Error bars show estimated one standard deviation.

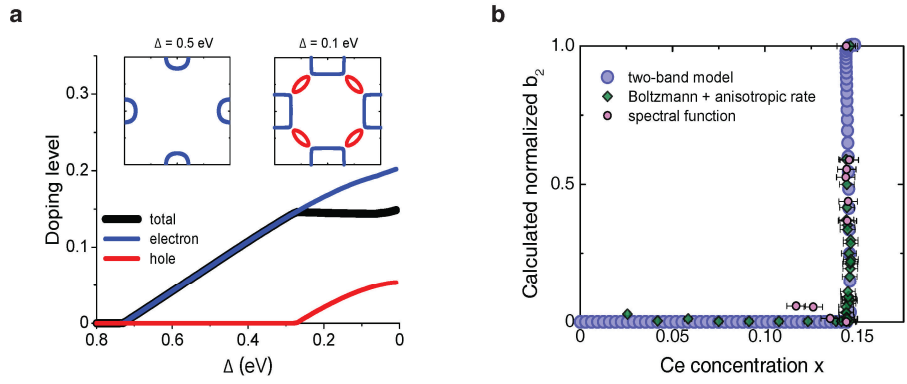

**Fig S3. Fermi surface and calculation of magnetoresistivity.** (a) FS and carrier densities calculated based on the local density approximation[10,11], with the constraint $n_e - n_h = x$, where $x$ is the Ce concentration. Hole Fermi pockets appear at $x \approx 0.145$. (b) MR calculated using three methods.

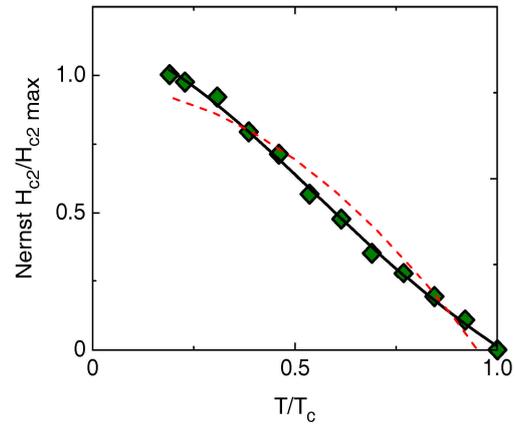

**Fig S4. Nernst upper critical field**. Green diamonds are data adopted from ref. 37. Dashed red and solid black lines are best fits to single- and two-band model, respectively.

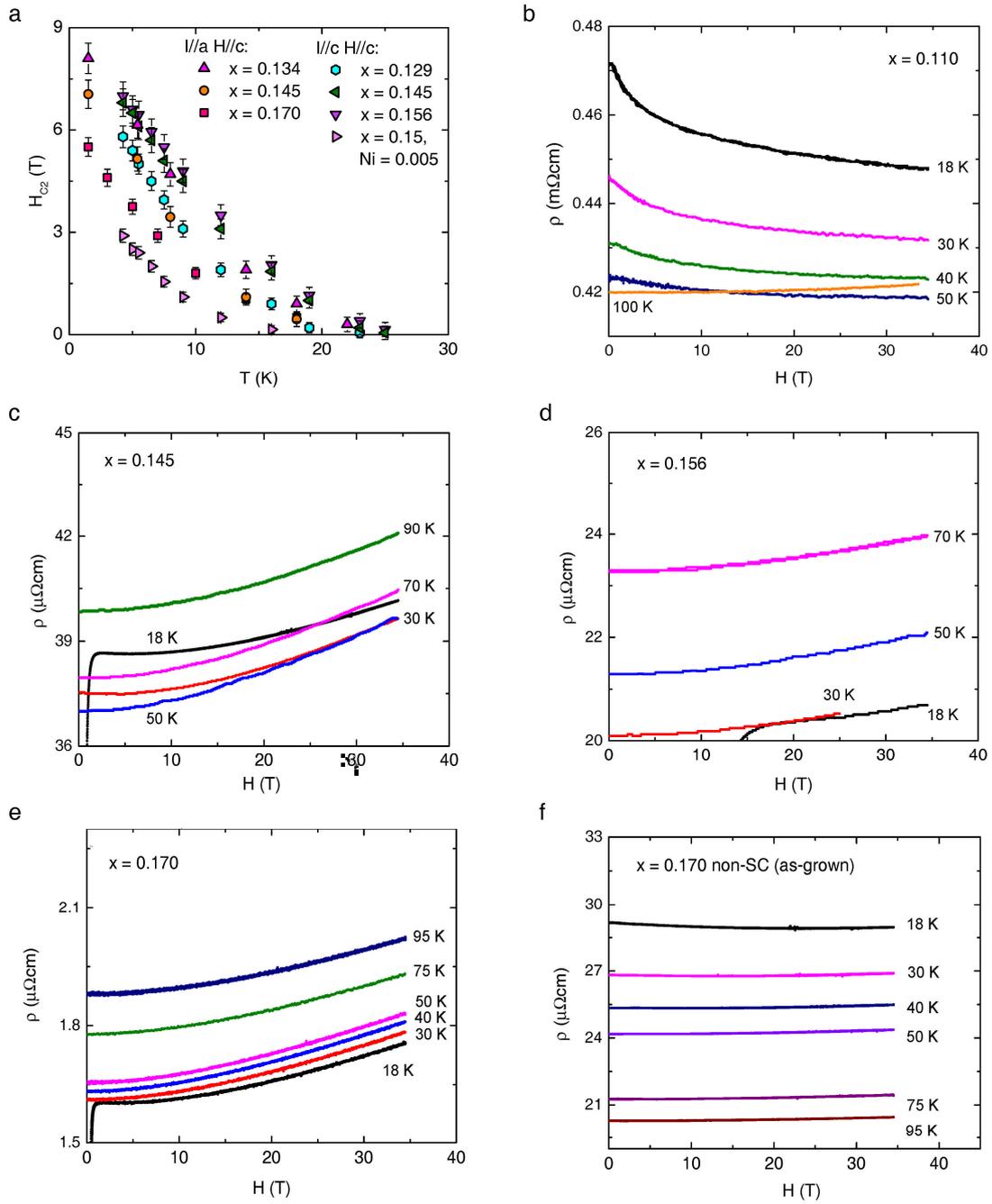

**Fig S5. Representative MR and $H_{C2}$ data**. **(a)** Unscaled $H_{C2}$ for NCCO at various chemical composition and temperatures. **(b-f)** Representative raw MR data for NCCO at various chemical composition and temperatures.